\documentclass{PoS}

\def\siml{{\ \lower-1.2pt\vbox{\hbox{\rlap{$<$}\lower6pt\vbox{\hbox{$\sim$}}}}\ }}
\def\simg{{\ \lower-1.2pt\vbox{\hbox{\rlap{$>$}\lower6pt\vbox{\hbox{$\sim$}}}}\ }}

\def \als {\alpha_{\mathrm{s}}}

\def \m2   {\mu^{2 \epsilon}}
\def\lQ{\Lambda_{\rm QCD}}
\def\als{\alpha_{\rm s}} 
\def\bea{\begin{eqnarray}}
\def\eea{\end{eqnarray}}
\def\be{\begin{equation}}
\def\ee{\end{equation}}

\title{Recent Developments in the Theory of Quarkonium}

\ShortTitle{HQL 2010}

\author{\speaker{Nora Brambilla}\\
        Physik-Department, Technische Universit\"at M\"unchen,
James-Franck-Str. 1, 85748 Garching, Germany\\
        E-mail: \email{nora.brambilla@ph.tum.de}}

\abstract{I review some recent developments in the theory of heavy quarkonium 
 in an effective field theory framework.}

\FullConference{The Xth Nicola Cabibbo International Conference on Heavy Quarks and Leptons,\\
		October 11-15, 2010\\
		Frascati (Rome) Italy}

\begin{document}

\section{Quarkonium}

In the last few years the B-factories Babar at SLAC and Belle at KEK, 
CLEO-III and Cleo-c at CESR,  CDF and D0 at  Fermilab and BESII and BESIII at IHEP 
have largely  operated as quarkonium factories producing large data sample on quarkonium
spectra,  decays  and production  with high statistics.
New states, new production mechanisms, new transitions and  unexpected states of an exotic nature 
have been observed. The study of quarkonium in media has also undergone crucial developments,
the suppression of quarkonium production in the hot medium remaining one of  the 
cleanest and most relevant probe of deconfined matter.
More is expected to come soon from LHC experiments.

From the theory point of view,  effective field theories  (EFTs)  as HQEFT, NRQCD, pNRQCD, SCET...,  for the description of quarkonium processes have been newly developed and are being 
developed, providing a unifying description as well as a solid and versatile 
tool giving well-defined, model independent and precise predictions
\cite{Brambilla:2010cs,Brambilla:2004jw}.
They rely on one  hand on  high order perturbative calculations and on 
the other hand on lattice simulations, the recent  progress in both fields 
having added a lot to the theory reach.

The International Quarkonium Working Group  (QWG) (www.qwg.to.infn.it)
created in 2002 has supplied ad adequate platform for discussion and common researrch work between theorists 
and experimentalists, producing also two large reviews of  state of the art, open problems,  perspective and opportunities 
of quarkonium physics  in 2010 and 2004 \cite{Brambilla:2010cs}.

Here I will summarize some of the recent progress in theory.

\section{Effective Field Theories for Quarkonium}

The modern approach to quarkonium physics consists in taking  
advantage of the hierarchy of non-relativistic (NR) energy scales in the system 
by constructing a suitable hierarchy of effective field theories (EFTs)~\cite{Brambilla:2004jw}.

The energy scales are:  the heavy-quark mass (hard scale), 
$m$, the typical momentum transfer (soft scale), 
$p \sim mv$, whose inverse sets the typical distance, 
$r$, between the heavy quark and the antiquark, 
and the typical kinetic energy (ultrasoft scale), 
$E \sim mv^2$, whose inverse sets the typical time scale of the bound state. 
The heavy-quark bound-state velocity $v$ is a small quantity $v\ll 1$
($v^2 \sim 0.1$ for $b\bar{b}$, $v^2 \sim 0.3$ for $c\bar{c}$,  
$v^2 \sim 0.01$ for $t\bar{t}$), the mass is a large quantity $m\gg \lQ$,
$\als(m) \ll 1$.
For energy scales close to $\lQ$, perturbation theory breaks down  
and one has to rely on nonperturbative 
methods. Regardless of this, the nonrelativistic hierarchy of scales: 
$m \gg p \sim 1/r \sim mv  \gg E \sim m v^2$
also persists below the $\lQ$ threshold.
 While the hard scale is always  larger than 
$\lQ$, different  situations may arise for the other two scales 
depending on the considered quarkonium system.
The  soft scale, proportional to the  inverse typical radius $r$,
may be a perturbative ($\gg \lQ$) or a nonperturbative scale ($\sim \lQ$) depending on the physical system. 
The first case is likely to happen only for the lowest charmonium and 
bottomonium states. We do not have direct
information on the radius of the quarkonia systems, and thus the
attribution of  some of the lowest bottomonia and charmonia states 
to the perturbative or the nonperturbative soft regime is at the
moment still ambiguous.
The ultrasoft scale  may still be perturbative only in 
the case of  $t \bar{t}$ threshold states.
All quarkonium scales get entangled in a typical amplitude involving a quarkonium
observable. In particular, quarkonium annihilation 
and production take place at the scale $m$, quarkonium binding takes  place at the scale
$mv$, which is the typical momentum exchanged inside the bound state, while 
very low-energy gluons and light quarks (also called ultrasoft degrees of freedom)  
live long enough that a bound state has time to form and, therefore, are sensitive to the 
scale $mv^2$. Ultrasoft gluons are responsible for phenomena 
like the Lamb shift in QCD.

A hierarchy of EFTs may be constructed by systematically integrating out 
modes associated to  high energy scales not relevant for quarkonium.
Such integration  is made  in a matching procedure enforcing 
the  equivalence between QCD and the EFT at a given 
order of the expansion in $v$.
The EFT  realizes a factorization at the Lagrangian level between 
the high energy contributions, encoded into the  matching coefficients, and 
 the low energy contributions, carried by the dynamical degrees of freedom.
Poincar\'e symmetry remains  intact in a nonlinear realization at the level of the NR EFT
and imposes exact relations among the 
matching coefficients  \cite{Brambilla:2003nt}.

\section{Physics at the scale $m$: NRQCD}
At the scale $m$ the suitable EFT is NonRelativistic QCD (NRQCD)  \cite{Caswell:1985ui}, which 
follows from QCD by integrating out the scale $m$. As a consequence, 
the effective Lagrangian is organized as an expansion in $1/m$  and $\als(m)$.
The field of quarkonium production has seen terrific progress in the last 
few years both in theory and in experiments, for a review see  \cite{ Brambilla:2010cs,Bodwin:2010py}.

For what concerns decays,  recently, substantial progress has been made in the evaluation of the factorization formula at order $v^7$ \cite{Brambilla:2006ph}, in the lattice evaluation of the NRQCD matrix elements 
\cite{Bodwin:2005gg} and in the new,  accurate data on many hadronic 
and electromagnetic decays \cite{Brambilla:2010cs}. 
The data are clearly sensitive to NLO corrections in the Wilson coefficients 
and presumably also to relativistic corrections. Improved  theory predictability would entail  the lattice 
calculation or data extraction of the NRQCD matrix elements and 
perturbative resummation of large contribution in the NRQCD matching coefficients.
The new data on hadronic transitions and hadronic decays pose interesting 
challenging to the theory.

Using the new CLEO data on radiative $\Upsilon(1S)$ decay and the improved lattice determination of the NRQCD matrix elements
it has been possible to obtain in \cite{Brambilla:2007cz} a determination of $\als$ at the $\Upsilon$ mass 
 $\als(M_\Upsilon(1S))=0.184^{+0.015}_{-0.014}$  giving a value  $\als(M_z)=0.119^{0.006}_{-0.005}$ in agreement 
 with the world average.

Lattice NRQCD calculations have undergone a steady development in last few years see \cite{Brambilla:2010cs, Davies:2008hs}.

\section{Physics at the scale $mv$ and $mv^2$: pNRQCD}
At the scales $mv$ and $mv^2$ 
the suitable EFT is potential 
  NonRelativistic QCD (pNRQCD)   
\cite{Pineda:1997bj,Brambilla:1999xf}, which 
follows from NRQCD by integrating out the scale $mv$.

For quarkonium states away from threshold we have now a clear effective field 
description, based on perturbative and lattice computations.
This is nowadays the standard description.

The soft scale $mv$  may be larger or not than the confinement 
scale $\lQ$ depending on the radius of the quarkonium system. 
When $mv^2 \sim \lQ$, we speak about weakly-coupled pNRQCD because 
the soft scale is perturbative and the matching from NRQCD to pNRQCD 
may be performed in perturbation theory. 
When $mv  \sim \lQ$, we speak about  
strongly-coupled pNRQCD because the soft scale 
is nonperturbative and the matching from NRQCD to pNRQCD may not be performed in
perturbation theory \cite{Brambilla:2004jw}.

The potential is a Wilson coefficient  of the EFT. 
In general undergoes renormalization, develops scale dependence and satisfies renormalization 
group equations  which allow to resum large logarithms.  In this framework the $Q\bar{Q}$  potential, 
which is a fundamental object for QCD, can be defined and systematically calculated.

In the following we will summarize some recent phenomenological applications of pNRQCD.
There are several cases for the physics at hand. 
In the case in which the EFT has been constructed \cite{Brambilla:1999xf,Brambilla:2000gk,Brambilla:2004jw},
 i.e. for states below 
threshold,  the work is currently going in calculating higher 
order perturbative corrections in $v$ and $\als$ for processes of interest,
resumming the logarithms in the ratio of the scales that may be sizeable,
calculating or extracting nonperturbatively low energy correlators and extending the theory with the addictions of electromagnetic effects \cite{Brambilla:2005zw} and  the consideration 
of $QQQ$ and $QQq$ systems \cite{Brambilla:2009cd}.
The issue here is precision physics  and the study of confinement.
Close to threshold the full EFT has not yet been constructed and the degrees 
of freedom have still to be identified \cite{Brambilla:2008zz,Brambilla:2010cs}.
At finite temperature the EFT is being constructed and the existing results 
hint at a new physical picture with possible application at heavy ion 
collisions at LHC.

\subsubsection{Potentials, spectrum, decays for quarkonia of small radius}
If the quarkonium system is small, the soft scale is perturbative and the 
potentials can be  entirely  calculated in perturbation theory 
\cite{Brambilla:2004jw}.

Since the degrees of freedom that enter the Schr\"odinger description 
are in this case both $Q\bar{Q}$  color singlet and $Q\bar{Q}$ color octets,
both singlet and octet potentials exist.
The static singlet $Q \bar{Q}$ potential is pretty well known.
The three-loop correction to the static potential is now completely
known: the fermionic contributions to the three-loop 
coefficient~\cite{Smirnov:2008pn} first became available, and more
recently the remaining purely gluonic term has been 
obtained~\cite{Anzai:2009tm,Smirnov:2009fh}. 

The first log related to ultrasoft effects arises at three 
loops   \cite{Brambilla:1999qa} . Such logarithm  contribution at N$^3$LO 
and the single logarithm contribution at N$^4$LO may be extracted respectively 
from a one-loop and two-loop  calculation in the EFT and have been calculated 
in \cite{Brambilla:2010pp,Brambilla:2009bi}.

The perturbative series of the static potential suffers from a renormalon ambiguity 
(i.e. large $\beta_0$  contributions) and from large logarithmic contributions.
The singlet 
static energy,  given by the sum of a constant, the static potential and the ultrasoft 
corrections,
is free from ambiguities of the perturbative series. By resumming the large logs using 
the renormalization  group equations and  comparing it
(at the NNLL) with lattice 
calculations of the static energy one sees 
that the QCD perturbative series converges very nicely 
to and agrees with 
the lattice result in the short range    (up to 0.25~fm) and that no nonperturbative
linear (``stringy'') contribution to the static potential exist \cite{Pineda:2002se,Brambilla:2010pp}.

In particular, the 
recently obtained theoretical expression~\cite{Brambilla:2010pp} 
for the complete QCD static
energy at  NNNLL precision has
been used 
to determine $r_0 \Lambda_{\overline MS}$ by comparison with available lattice
data, where $r_0$ is the lattice scale and $\Lambda_{\overline MS}$
is the QCD scale, obtaining 
$r_0\Lambda_{\overline MS} =0.622^{+0.019}_{-0.015}$  for the zero-flavor case. 
This extraction was previously performed
at the NNLO level (including an estimate at NNNLO) in \cite{Sumino:2005cq}.
The same procedure can be used to obtain a precise evaluation of the
unquenched $r_0 \Lambda_{\overline MS}$ value after short distance unquenched
lattice data for the $Q \overline{Q}$ exist  \cite{Donnellan:2010mx}.

The static octet potential is known up to two loops~\cite{Kniehl:2004rk}.
Relativistic corrections to the static singlet potential
have been calculated over the years and are 
summarized in \cite{Brambilla:2004jw}. 

In the case of $QQq$ baryons, the static potential has been determined up to 
NNLO in perturbation theory    \cite{Brambilla:2009cd}   and recently also on the lattice \cite{Yamamoto:2007pf}.
Terms suppressed by powers of  $1/m$  and $r$ in the Lagrangian have been matched 
(mostly) at leading order and used to determine, for instance, the expected 
hyperfine splitting of the ground state of these systems.

In the case of $QQQ$ baryons, the static potential has been determined up to 
NNLO in perturbation theory \cite{Brambilla:2009cd}
and also on the lattice \cite{Takahashi:2000te}. The transition region from 
a Coulomb to a linearly raising potential is characterized in this case also 
by the emergence of a three-body potential apparently parameterized by only one length. 
It has been   shown that in 
perturbation theory a smooth genuine three-body potential shows up at two loops.

The energy levels have been calculated at order $m \als^5$  \cite{Brambilla:1999xj}.
Decays amplitude   \cite{Kiyo:2010jm,Brambilla:2010cs,Brambilla:2004jw}        and production and annihilation 
\cite{Beneke:2007pj} have been calculated in perturbation theory at high order.
An effective field theory of magnetic dipole transition has been given in 
\cite{Brambilla:2005zw} and a description of the $\eta_c$ line shape in \cite{Brambilla:2010ey}.

\subsubsection{Potentials, spectra and decays for quarkonia of large radius}
If the quarkonium system is large, the soft scale is nonperturbative and the 
potentials cannot be  entirely calculated in perturbation theory 
\cite{Brambilla:2004jw}.
Then the  potential matching coefficients
are obtained in the form of expectation values of gauge-invariant 
Wilson-loop operators. 
In this case, heavy-light meson pairs and heavy hybrids 
develop a mass gap of order $\lQ$ with respect to the energy of the
$Q\overline{Q}$ pair, the second circumstance 
being apparent from lattice simulations.
Thus, away from threshold, 
the quarkonium singlet field $S$ is the only low-energy dynamical 
degree of freedom in the pNRQCD Lagrangian 
(neglecting ultrasoft corrections coming 
from pions and other Goldstone bosons).
The singlet potential $V_S(r)$ can be expanded
in powers of the inverse of the quark mass;
static, $1/m$ and $1/m^2$ terms were calculated long 
ago~\cite{Brambilla:2000gk}.
They involve NRQCD matching coefficients (containing 
the contribution from the hard scale) and low-energy 
nonperturbative parts given in terms
of static Wilson loops and field-strength insertions in the static
Wilson loop
(containing the contribution from the soft scale).
Such expressions correct and generalize previous finding in the Wilson loop approach
\cite{Eichten:1980mw} 
that were typically missing the high energy parts of the potentials, 
encoded into the NRQCD matching coefficients and containing the 
dependence on the logarithms of $m_Q$, and some of the low energy contributions.

In this regime of pNRQCD, we recover the quark potential singlet model. 
However, here the potentials are calculated in QCD by nonperturbative 
matching. Their evaluation requires calculations on the lattice 
or in QCD vacuum models \cite{Brambilla:1999ja}.
Recent progress includes new, precise lattice calculations 
of these potentials 
obtained using the L\"uscher multi-level algorithm \cite{Koma:2007jq}.
The nonperturbative potentials for the $QQQ$ and $QQq$  have been obtained in 
 the second reference of  \cite{Brambilla:2009cd}  and in \cite{Brambilla:1993zw}.
 Inclusive decay amplitudes have been calculated (see the last ref. in \cite{Brambilla:2000gk})
 and the number of nonperturbative correlators appears to be sizeably reduced with respect to NRQCD.

\subsection{Quarkonium potential at finite T and heavy ion collisions}

In the last few years years, there has been a remarkable 
progress in constructing EFTs  for quarkonium at finite temperature and 
in rigorously defining  the quarkonium potential \cite{finite,Brambilla:2008cx}.
 Quarkonium in a medium is characterized by different energy and momentum scales;  there are the scales of the non-relativistic bound state  
 and there are the  
thermodynamical scales: the temperature $T$, the inverse of the screening 
length of the chromoelectric interactions, i.e. the Debye mass $m_D$ and lower scales.
Up to now calculations are done in the weak coupling regime.  Integrating out 
sequentially the energy scales a version of pNRQCD at finite T has been obtained 
and the potential has been calculated as matching coefficient of the EFT.
Such potential has new and unexpected feautures, e.g. a large imaginary contribution,
 that will have an important impact in phenomenological studies of quarkonium 
 suppression at RHIC and at LHC. In particular in  \cite{Brambilla:2010vq} 
heavy quarkonium energy levels and decay widths in a quark-gluon plasma, 
below the melting temperature at a
temperature T and screening mass $m_D$ satisfying the hierarchy  
$m \als  \gg \pi T  \gg m \als^2 \gg m_D$, have been calculated  at order $m \als^5$.
This  situation is relevant for bottomonium $1S$   states ($\Upsilon(1S)$, $\eta_b$) at the LHC.


\end{document}